\documentclass[journal=apchd5,manuscript=article]{achemso} %ancac3, acbcct, apchd5
\setkeys{acs}{keywords = true}

\usepackage[version=3]{mhchem} % Formula subscripts using \ce{}
\usepackage{bm} 
\usepackage{amssymb}
\def\gsim{\mathrel{\rlap{\lower4pt\hbox{\hskip1pt$\sim$}}
    \raise1pt\hbox{$>$}}}    
\author{Anders Pors}
\email{alp@iti.sdu.dk}
\author{Sergey I. Bozhevolnyi}
\affiliation{Department of Technology and Innovation, University of Southern Denmark, Niels Bohrs All{\'{e}} 1, DK-5230 Odense M, Denmark}

\title[Emitters near Layered Plasmonic Nanostructures]
{Quantum Emitters near Layered Plasmonic Nanostructures: Decay Rate Contributions}


\begin{document}
%%%%%%%%%%%%%%%%%%%%%%%%%%%%%%%%%%%%%%%%%%%%%%%%%%%%%%%%%%%%%%%%%%%%%
%% The manuscript does not need to include \maketitle, which is
%% executed automatically.  The document should begin with an
%% abstract, if appropriate.  If one is given and should not be, the
%% contents will be gobbled.
%%%%%%%%%%%%%%%%%%%%%%%%%%%%%%%%%%%%%%%%%%%%%%%%%%%%%%%%%%%%%%%%%%%%%
\begin{abstract}
We introduce a numerical framework for calculating decay rate contributions when excited two-level quantum emitters are located near layered plasmonic nanostructures, particularly emphasizing the case of plasmonic nanostructures atop metal substrates where three decay channels exist: free space radiation, Ohmic losses, and excitation of surface plasmon polaritons (SPPs). The calculation of decay rate contributions is based on Huygen's equivalence principle together with a near-field to far-field transformation of the local electric field, thereby allowing us to discern the part of the electromagnetic field associated with free propagating waves rather than SPPs. The methodology is applied to the case of an emitter inside and near a gap-plasmon resonator, emphasizing strong position and orientation dependencies of the total decay rate, contributions of different decay channels, radiation patterns, and directivity of SPP excitation.
\end{abstract}

%\section{Introduction}
Ever since the pioneering work of E. M. Purcell\cite{purcell}, stating that spontaneous emission from quantum emitters (QEs) is not an intrinsic property but can be modified by introducing inhomogeneities in the nearby surroundings, researchers and engineers have studied a multitude of systems, including planar interfaces\cite{lukosz,ford}, cavities\cite{bjork,gerard}, photonic crystals\cite{yablonovitch,petrov}, waveguides\cite{kleppner,chen,shailesh,shailesh2}, and plasmonic (i.e., metallic) nanostructures\cite{anger,kuhn,kinkhabwala}. Especially the interaction of QEs with metallic objects, being either waveguiding or finite-sized structures, has shown the possibility to enhance the decay rate by several orders of magnitude\cite{ford,chen,kinkhabwala} due to the strong confinement of the electromagnetic field at metal-dielectric interfaces. Nevertheless, plasmonic structures support a high number of dissipative states that are probed by the near-field of a QE, leading, despite a strongly enhanced decay rate, to emission quenching (or at least low quantum yield) due to domination of non-radiative decay channels. Since a high rate of spontaneous emission can improve efficiency of certain optoelectronic devices, such as light emitting diodes\cite{fan} and single-photon sources\cite{eisaman}, it is at the heart of current research to investigate configurations with high field enhancement and reasonable ratio between radiative and non-radiative decay probabilities. One such promising geometry consists of a metal film overlaid with a QE-doped nanometer-thin dielectric layer supporting rectangular or circular metal nanoparticles\cite{russell,rose,ciraci,akselrod}, hence featuring gap-surface plasmon (GSP) resonances and, for this reason, is also known as GSP resonators\cite{nielsen}. It should, however, be noted that the proximity of the QE and the metal film facilitates the excitation of surface plasmon polaritons (SPPs) which, in addition to the radiative and non-radiative relaxation paths, can be considered as a third decay channel. In the quest for high radiative decay rates, one typically does not differentiate between SPP and non-radiative decay channels, but we would like to emphasize that in certain cases, such as the design of efficient local sources for plasmonic circuitry\cite{chen,shailesh,shailesh2}, it is of great importance to know the rate of SPP\cite{ford} or waveguide mode\cite{chen} excitation. For this reason, we propose a numerical methodology, based on Huygen's equivalence principle and a near-field to far-field (NF2FF) transformation of the local electric field\cite{michalski,capoglu,muller}, that, unlike other numerical studies\cite{ciraci}, allow for accurate calculations of the radiative, non-radiative, \emph{and} SPP decay rates for layered plasmonic systems with arbitrary-shaped inclusions. The calculation procedure is verified for QEs above a metal film and applied to the study of GSP resonators, emphasizing strong position and orientation dependencies of QE of the total decay rate, influence of decay channels, radiation patterns, and directionality of SPP excitation -- features that all can be important depending on the application.

\section{RESULTS AND DISCUSSION}

\subsection{Basic Equations}%Decay rate contributions}
In the regime of weak coupling between light and matter, the effect of an electromagnetic field acting on an QE can be described perturbatively, meaning that the light field modifies only the decay rate of spontaneous emission. Moreover, it is generally accepted that the relative change in the spontaneous decay rate of a two-level QE (i.e., the Purcell factor), $\gamma_{\mathrm{tot}}/\gamma_0$, where $\gamma_{0}$ is the free space decay rate and $\gamma_{\mathrm{tot}}$ is the modified rate due to inhomogeneities in the nearby surroundings, can be calculated using classical calculation of the normalized power dissipated by an electric dipole, $P_{\mathrm{tot}}/P_0$, in this inhomogeneous environment\cite{novotny}:
\begin{equation}
\frac{\gamma_{\mathrm{tot}}}{\gamma_0}=\frac{\omega}{2P_0}\mathrm{Im}\left\{\bm{\mu}^*\cdot\mathbf{E}(\mathbf{r}_0)\right\}.
\label{eq:gammatot}
\end{equation}
Here, $\omega$ is the angular frequency, $P_0=|\bm{\mu}|^2\omega\sqrt{\varepsilon}k_0^3/(12\pi\varepsilon_0)$ is the power radiated in homogeneous space with relative permittivity $\varepsilon$, $k_0$ is the vacuum wave number, $\varepsilon_0=8.854\cdot10^{-12}$\,F/m is the vacuum permittivity, $\bm{\mu}$ is the dipole moment, $^*$ means complex conjugate, and \textbf{E} is the electric field evaluated at the position of the dipole $\mathbf{r}_0$.

It should be noted that an excited QE in inhomogeneous surroundings not only relaxes to the ground state via spontaneous emission of photons, but may also decay by Ohmic heating in lossy media and/or by excitation of surface and waveguide modes in layered geometries\cite{novotny}. In this work, we consider a layered geometry, as shown in Figure \ref{fig:Sketch3LayerSystem}, in which the lower medium (medium 2) is metal and may be decorated by a subwavelength-thick spacer layer (medium 3) with the QE being in close vicinity of an arbitrarily-shaped metal nanostructure. As such, the total spontaneous decay rate can be written as $\gamma_{\mathrm{tot}}=\gamma_{\mathrm{rad}}+\gamma_{\mathrm{spp}}+\gamma_{\mathrm{abs}}$, where $\gamma_{\mathrm{rad}}$, $\gamma_{\mathrm{spp}}$, and $\gamma_{\mathrm{abs}}$ are the decay rates into free space radiation, excitation of surface plasmon polaritons (SPPs), and absorption in metal, respectively. 
% Figure 1
The normalized radiative decay rate can be found by integrating the time-averaged far-field Poynting vector on a hemisphere in the upper dielectric medium (medium 1) and dividing by $P_0$, i.e., 
\begin{equation}
\frac{\gamma_{\mathrm{rad}}}{\gamma_0}=\frac{1}{2P_0}\sqrt{\frac{\varepsilon_0\varepsilon}{\mu_0}}\iint_{\mathrm{HS}}{|\mathbf{E}_{\mathrm{ff}}|^2}dS,
\label{eq:gammarad}
\end{equation}   
where $\mathbf{E}_{\mathrm{ff}}$ is the electric far-field. Secondly, we can calculate the SPP decay rate by realizing that the difference in power leaving the surface $\Gamma$ in the near-field region of the QE (see Figure \ref{fig:Sketch3LayerSystem}) and the power reaching the far-field must be equal to the power carried by SPPs. Consequently, 
\begin{equation}
\frac{\gamma_{\mathrm{spp}}}{\gamma_0}=\frac{1}{P_0}\iint_\Gamma{\mathbf{S}\cdot\hat{\mathbf{n}}}\,dS-\frac{\gamma_{\mathrm{rad}}}{\gamma_0},
\label{eq:gammaspp}
\end{equation}
where $\mathbf{S}=1/2\mathrm{Re}\left\{\mathbf{E}\times\mathbf{H}^*\right\}$ is the time-averaged Poynting vector, $\mathbf{H}$ is the magnetic field, and $\hat{\mathbf{n}}$ is the outward-pointing normal vector to $\Gamma$. Finally, the nonradiative decay channel can be quantified by the following relation 
\begin{equation}
\frac{\gamma_{\mathrm{abs}}}{\gamma_0}=\frac{\gamma_{\mathrm{tot}}-\gamma_{\mathrm{rad}}-\gamma_{\mathrm{spp}}}{\gamma_0},
\label{eq:gammaabs}
\end{equation} 
where the quantities in the numerator follow from eqs \ref{eq:gammatot}-\ref{eq:gammaspp}.

Having setup the basic equations used in this work, we would like to add a few comments to the assumptions and approximations involved. Firstly, it should be noted that the validity of eq \ref{eq:gammaspp} implies the assumption of subwavelength-thin dielectric spacer so that it does not support photonic waveguide modes. Secondly, since SPPs attenuate as they propagate along the metal surface, we need to put restrictions on the dimension of $\Gamma$ in order to ensure the correct balance between rate of SPP generation and quenching. As the non-radiative contribution to the total decay rate can be viewed as excitation of lossy surface waves that decay within a fraction of a wavelength\cite{ford}, we define the lower bound $\lambda/4<L_\Gamma$, with $L_\Gamma$ being the distance from QE to $\Gamma$. The upper bound on $L_\Gamma$, on the other hand, is limited by the propagation length of the SPP, $L_{\text{spp}}$, hereby leading to the condition $L_\Gamma\ll L_{\text{spp}}$. In the following sections, we investigate QE emission at the wavelength $\lambda=780$\,nm for which $\lambda \ll L_{\text{spp}}$, thus resulting in a proper balance between decay rate contributions when $L_\Gamma\sim\lambda/4-\lambda$. As a study example, Figure S1 in Supporting Information (SI) demonstrates the weak dependence of decay rates on $L_\Gamma$ for a vertically-oriented QE positioned in air and 20\,nm above a gold substrate. In contrast, Figure S2 in SI displays for the same configuration the dependence of decay rates as a function of separation when emission occurs at $\lambda=500$\,nm and $\lambda=550$\,nm in which $L_\Gamma \gsim L_{\text{spp}}$, hereby leading to an improper balance between the absorptive and SPP decay rates. 
Additionally, it is worth noting that the calculation of the absorptive decay rate using eq \ref{eq:gammaabs} is a computationally inexpensive calculation that automatically ensures the correct energy balance. However, as discussed by C. Cirac{\'i} \emph{et al.}\cite{ciraci}, one can also calculate $\gamma_{\mathrm{abs}}/\gamma_0$ directly by integrating the Ohmic heating density [$U_h(\omega,\mathbf{r})=\tfrac{1}{2}\omega\varepsilon_0\varepsilon_m^{''}|\mathbf{E}|^2$, where $\varepsilon_m^{''}$ is the imaginary part of the metal permittivity and $\mathbf{E}$ is the electric field in the metal] in the metal bounded by $\Gamma$. As such, the comparison of eq \ref{eq:gammaabs} with the direct calculation of $\gamma_{\mathrm{abs}}/\gamma_0$ can be used to benchmark the numerical accuracy of the calculations devised in eqs \ref{eq:gammatot}--\ref{eq:gammaabs}. As an example, Figure S3 in SI displays the absorptive decay rate obtained by both mentioned procedures for a vertical QE in air positioned above a gold substrate. It is seen that the two procedures give almost identical results when quenching constitutes an appreciable fraction of the total decay rate. 
As a final comment, it ought to be mentioned that for an optically-thick spacer, in which plasmonic and photonic propagating modes coexist, the SPP decay rate (eq \ref{eq:gammaspp}) turns into a total waveguide mode decay rate that cannot distinguish the contributions from the different propagating modes. We envision that the relative contributions of the different modes can be obtained by projecting the total electric field on the different mode fields.

\subsection{Near-Field to Far-Field Transformation}
In the study of QEs near layered plasmonic nanostructures and the coupling to different decay channels, it is clear from the previous section that one must know the electric far-field (see eq \ref{eq:gammarad}). Although some numerical (or semi-analytical) approaches allow for a direct evaluation of the far-field, such as, e.g., the boundary element method\cite{jung}, the finite element approach (FEA) utilized in the work permits (for computational reasons) only to evaluate the electromagnetic field in a limited space around the dipole source. The advantage of FEA is the broad applicability and the superior meshing capabilities, allowing one to study practically any kind of geometries and material compositions, but in order to obtain the far-field one must employ a NF2FF transformation\cite{jin}. Such NF2FF transformations all rely on Huygens equivalence principle and knowledge of Green's functions for the reference geometry (i.e., without sources and nanostructures)\cite{michalski,capoglu,muller}. The equivalence principle states that the respective electric and magnetic surface currents 
\begin{equation}
\mathbf{J}_s=\hat{\mathbf{n}}\times\mathbf{H} \ \ \ ,\ \ \ \mathbf{M}_s=-\hat{\mathbf{n}}\times\mathbf{E},
\label{eq:sc}
\end{equation}
defined on a closed fictitious surface surrounding all sources and scatterers, create the same field outside of the surface as the original problem, just with the geometry being the simpler reference geometry. In our case, with the lower medium being metal, we approximate the closed surface with an open surface $\Gamma$ (see Figure \ref{fig:Sketch3LayerSystem}) and the reference geometry is the three-layered media with interfaces at $z=0$ and $z=-t_s$. It should be stressed that that usage of an open surface is only exactly valid for a perfect metal\cite{capoglu}, since in this case no field (i.e., Huygen sources) exists in the metal. Noting, however, that the electromagnetic field is strongly attenuated within few tenth of nanometers inside a good conductor, it is clear that this field, constituting Huygen sources, will be strongly attenuated before it reaches back to the metal-dielectric interface where it can be transmitted and, hence, contribute to the far-field response. As an example of the good approximation involved in choosing an open surface $\Gamma$, Figure S4 in SI displays the almost identical radiation patterns of a vertical dipole above a gold substrate when $\Gamma$ is chosen as an open and closed surface, respectively. In this regard, we would like to point out that for configurations containing optically-thin metal films, it is in general advisable to use closed fictitious surfaces.

Assuming for the moment that the Green's dyadics for the reference geometry are constructed, the electric field at position \textbf{r} outside of $\Gamma$ is represented by the expression [time convention: $\exp(-i\omega t)$]
\begin{equation}
\mathbf{E}(\mathbf{r})=i\omega\mu_0\iint_\Gamma\stackrel{\leftrightarrow}{\mathbf{G}}_J(\mathbf{r},\mathbf{r}')\mathbf{J}_s(\mathbf{r}')\,dS' +i\omega\varepsilon_0\iint_\Gamma\stackrel{\leftrightarrow}{\mathbf{G}}_M(\mathbf{r},\mathbf{r}')\mathbf{M}_s(\mathbf{r}')\,dS',
\label{eq:E}
\end{equation}
where $\stackrel{\leftrightarrow}{\mathbf{G}}_J$ and $\stackrel{\leftrightarrow}{\mathbf{G}}_M$ are the electric and magnetic Green's dyadics, respectively. In evaluating the electric far-field, however, one can take advantage of the translational invariance of the reference geometry in the $xy$-plane, hereby allowing for a 2D Fourier transform of the electric field in a plane $z>0$, i.e. %thus defining the 2D Fourier spectrum of the electric field in a plane $z>0$ as
\begin{align}
\tilde{\mathbf{E}}(\mathbf{K},z)=&i\omega\mu_0\iint_\Gamma\stackrel{\leftrightarrow}{\tilde{\mathbf{G}}}_J(\mathbf{K},z,z')\mathbf{J}_s(\mathbf{r}')e^{-i\mathbf{K}\cdot\mathbf{R}'}\,dS'+\nonumber \\
&i\omega\varepsilon_0\iint_\Gamma\stackrel{\leftrightarrow}{\tilde{\mathbf{G}}}_M(\mathbf{K},z,z')\mathbf{M}_s(\mathbf{r}')e^{-i\mathbf{K}\cdot\mathbf{R}'}\,dS',
\label{eq:E2}
\end{align}
with $\stackrel{\leftrightarrow}{\tilde{\mathbf{G}}}_J$ and $\stackrel{\leftrightarrow}{\tilde{\mathbf{G}}}_M$ being the angular spectrum representation of Green's dyadics, $\mathbf{K}=(k_x,k_y)$ is the in-plane wave vector, and $\mathbf{R}'=(x',y')$ is the in-plane position vector on the source surface $\Gamma$. As the electric field at each point in the far-field represents a plane wave with a specific wave vector (a fact utilized in eq \ref{eq:gammarad}), it naturally follows that the Fourier spectrum entirely defines the far-field. A thorough derivation shows that\cite{novotny}
\begin{equation}
\mathbf{E}_{\mathrm{ff}}(\hat{\mathbf{a}})=-2\pi ik_{1z} \tilde{\mathbf{E}}(\mathbf{K},z)e^{-ik_{1z}z}\frac{e^{ik_1r}}{r},
\label{eq:Efar}
\end{equation} 
where $\hat{\mathbf{a}}=\mathbf{r}/r$ is a unit vector pointing in the direction of observation, $r$ is the distance from origin to observation point, $k_1=k_0\sqrt{\varepsilon_1}$, and $k_{1z}=\sqrt{k_0^2\varepsilon_1-|\mathbf{K}|^2}$. 

The angular spectrum representation of Green's dyadics for layered geometries can be constructed in several ways, but one intuitive and elegant approach is developed by J. E. Sipe\cite{sipe}. Without dwelling on the details, the method immediately splits the fields generated by electric and magnetic sources into s- and p-polarized waves from which the interaction with material interfaces can be described by Fresnel transmission and reflection coefficients. In order to describe the propagation and polarization of plane waves, two right-handed triads $(\hat{\mathbf{s}},\hat{\mathbf{p}}_{i+},\hat{\mathbf{k}}_{i+})$ and $(\hat{\mathbf{s}},\hat{\mathbf{p}}_{i-},\hat{\mathbf{k}}_{i-})$ are defined (see Figure \ref{fig:Sketch3LayerSystem}), which describe upward and downward propagating waves in medium $i$, respectively, and can be expressed as
\begin{equation}
\hat{\mathbf{s}}=\hat{\mathbf{K}}\times \hat{\mathbf{z}} \ , \ \hat{\mathbf{p}}_{i\pm}=k_i^{-1}\left(K\hat{\mathbf{z}}\mp k_{iz}\hat{\mathbf{K}}\right) \ , \ \hat{\mathbf{k}}_{i\pm}=k_i^{-1}\left(K\hat{\mathbf{K}}\pm k_{iz}\hat{\mathbf{z}}\right),
\label{eq:triads}
\end{equation}
where $\hat{\mathbf{K}}=\mathbf{K}/K$, $K=|\mathbf{K}|$, and $k_i=k_0\sqrt{\varepsilon_i}$. The use of polarization vectors $\hat{\mathbf{s}}$ and $\hat{\mathbf{p}}_{i\pm}$ result in Green dyadics whose construction allow for an immediate verification by physical intuition. For example, if we at first consider a single interface at $z=0$ between metal and dielectric half-spaces (i.e., $\varepsilon_3=\varepsilon_2$ in Figure \ref{fig:Sketch3LayerSystem} and $\Gamma$ is limited to medium 1), the Green's dyadics for upward propagating waves in medium 1 take the form
\begin{align} \label{eq:GreenJ1layer}
\stackrel{\hspace{0.4mm}\leftrightarrow \hfill}{\tilde{\mathbf{G}}^{(1)}_J}(\mathbf{K},z,z')&=\frac{1}{8\pi^2k_{1z}}\left[\left(\hat{\mathbf{s}}\hat{\mathbf{s}} +\hat{\mathbf{p}}_{1+}\hat{\mathbf{p}}_{1+}\right)e^{ik_{1z}(z-z')}+\left(\hat{\mathbf{s}}r_{12}^s\hat{\mathbf{s}} +\hat{\mathbf{p}}_{1+}r_{12}^p\hat{\mathbf{p}}_{1-}\right)e^{ik_{1z}(z+z')}\right] \\
\stackrel{\hspace{0.4mm}\leftrightarrow \hfill}{\tilde{\mathbf{G}}^{(1)}_M}(\mathbf{K},z,z')&=\frac{\eta_0\sqrt{\varepsilon_1}}{8\pi^2k_{1z}}\left[\left(\hat{\mathbf{p}}_{1+}\hat{\mathbf{s}} -\hat{\mathbf{s}}\hat{\mathbf{p}}_{1+}\right)e^{ik_{1z}(z-z')}+\left(\hat{\mathbf{p}}_{1+}r_{12}^p\hat{\mathbf{s}} -\hat{\mathbf{s}}r_{12}^s\hat{\mathbf{p}}_{1-}\right)e^{ik_{1z}(z+z')}\right], 
\label{eq:GreenM1layer}
\end{align}
where $\eta_0=\sqrt{\mu_0/\varepsilon_0}$ is the vacuum wave impedance. In the equations above, the first term in square brackets denotes the direct upward propagation of s- and p-polarized plane waves to the observation plane at $z$, while the second term describes initially downward propagating waves (generated from sources on $\Gamma$) that reflect on the interface with amplitude and phase defined by the Fresnel reflection coefficients between medium $i$ and $j$
\begin{equation}
r_{ij}^s=\frac{k_{iz}-k_{jz}}{k_{iz}+k_{jz}} \ , \ r_{ij}^p=\frac{\varepsilon_j k_{iz}-\varepsilon_ik_{jz}}{\varepsilon_j k_{iz}+\varepsilon_ik_{jz}}.
\label{eq:Fresnell}
\end{equation}
Furthermore, note that the factors $\exp[ik_{1z}(z\mp z')]$ in eqs \ref{eq:GreenJ1layer} and \ref{eq:GreenM1layer} describe phase accumulation between source point $z'$ and observation plane $z$ $(\gg z')$ for the direct and reflected part of the electric field. 

Returning to the 3-layer geometry in Figure \ref{fig:Sketch3LayerSystem}, the problem becomes slightly more complicated since different Green's dyadics must be used for current sources in medium 1 and 3. Noting that the electric field generated by sources in medium 1 must, similar to the single interface case, consist of a direct and reflected part, the appropriate Green's dyadics correspond to eqs \ref{eq:GreenJ1layer} and \ref{eq:GreenM1layer} with the reflection coefficients replaced by the generalized counterpart for the 3-layer system
\begin{equation}
R_{12}^m=r_{13}^m+\frac{t_{13}^mr_{32}^mt_{31}^me^{2ik_{3z}t_s}}{1-r_{31}^mr_{32}^me^{2ik_{3z}t_s}},
\label{eq:r3layer}
\end{equation}
where the superscript $m$ denotes either s- or p-polarized light. The first term describes reflection at the interface between medium 1 and 3, while the second term accounts for downward propagating plane waves that transmit into medium 3, reflect at the interface between medium 3 and 2, and retransmit into medium 1, with the possibility of experiencing multiple reflections in medium 3 (as seen by the geometric series $(1-q)^{-1}=1+q+q^2+\cdots$). Regarding upward propagating waves in medium 1 emanating from sources in medium 3, the Green's dyadics must consist of contributions from initially upward propagating plane waves that are transmitted into medium 1, and initially downward propagating waves that reflect at the interface between medium 3 and 2, followed by transmission into medium 1. In both cases, however, light will undergo multiple reflections in medium 3, meaning that the appropriate Green's dyadics are of the form
\begin{align} \label{eq:GreenJ2layer}
\stackrel{\hspace{0.4mm}\leftrightarrow \hfill}{\tilde{\mathbf{G}}^{(3)}_J}(\mathbf{K},z,z')=&\frac{1}{8\pi^2k_{3z}}\Big[\left(\hat{\mathbf{s}}T_{31}^s\hat{\mathbf{s}} +\hat{\mathbf{p}}_{1+}T_{31}^p\hat{\mathbf{p}}_{3+}\right)e^{i(k_{1z}z-k_{3z}z')}+ \nonumber \\
& \left(\hat{\mathbf{s}}T_{31}^sr_{32}^s\hat{\mathbf{s}} +\hat{\mathbf{p}}_{1+}T_{31}^pr_{32}^p\hat{\mathbf{p}}_{3-}\right)e^{i(k_{1z}z+k_{3z}(2t_s+z'))}\Big], \\
\stackrel{\hspace{0.4mm}\leftrightarrow \hfill}{\tilde{\mathbf{G}}^{(3)}_M}(\mathbf{K},z,z')=&\frac{\eta_0\sqrt{\varepsilon_3}}{8\pi^2k_{3z}}\Big[\left(\hat{\mathbf{p}}_{1+}T_{31}^p\hat{\mathbf{s}} -\hat{\mathbf{s}}T_{31}^s\hat{\mathbf{p}}_{3+}\right)e^{i(k_{1z}z-k_{3z}z')}+
\nonumber \\
&\left(\hat{\mathbf{p}}_{1+}T_{31}^pr_{32}^p\hat{\mathbf{s}} -\hat{\mathbf{s}}T_{31}^sr_{32}^s\hat{\mathbf{p}}_{3-}\right)e^{i(k_{1z}z+k_{3z}(2t_s+z'))}\Big], 
\label{eq:GreenM2layer}
\end{align}
where
\begin{equation}
T_{31}^m=\frac{t_{31}^m}{1-r_{31}^mr_{32}^me^{2ik_{3z}t_s}},
\label{eq:t3layer}
\end{equation}
represents the generalized transmission coefficient from medium 3 to 1. Summarizing, we are now able to calculate the electric far-field in medium 1 for the three-layer geometry in Figure \ref{fig:Sketch3LayerSystem} by utilizing eqs \ref{eq:E2} and \ref{eq:Efar}, with source currents defined in eq \ref{eq:sc} and Green's dyadics in eqs \ref{eq:GreenJ1layer}, \ref{eq:GreenM1layer}, \ref{eq:GreenJ2layer}, and \ref{eq:GreenM2layer}.

\subsection{Quantum Emitter near Planar Metal Interface}
\begin{figure}[H]
	\centering
		\includegraphics[width=0.5\linewidth]{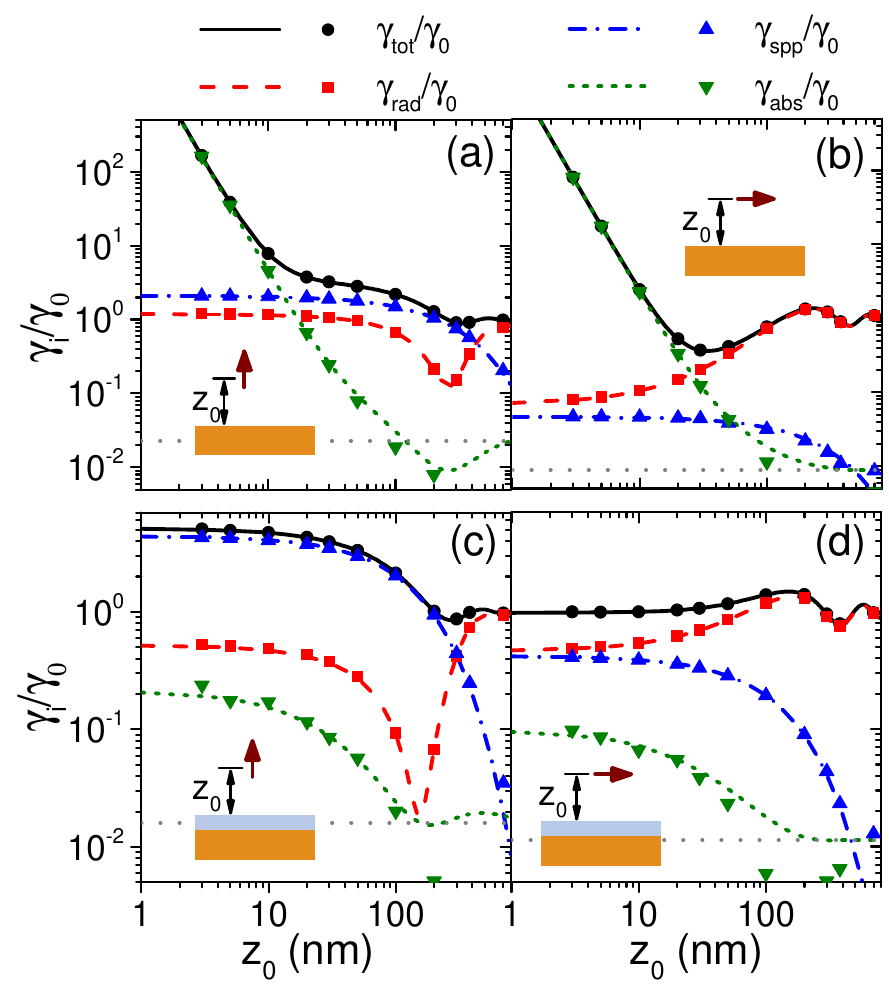}
	\caption{Relative decay rates for (a,c) vertical and (b,d) horizontal QE in air as a function of separation $z_0$ from (a,b) gold substrate and (c,d) gold substrate overlaid with 50\,nm of silicon dioxide. Markers correspond to numerical calculations (based on eqs \ref{eq:gammatot}-\ref{eq:gammaabs}) and lines are analytical curves. Gray dashed-dotted lines define the asymptotic value of $\gamma_{\mathrm{abs}}/\gamma_0$ for $z_0\rightarrow \infty$. The emission wavelength is in all cases $\lambda=780$\,nm.}
	\label{fig:DipolesLayeredGeom}
\end{figure}

As a way of benchmarking the proposed methodology for quantifying decay rate contributions of QEs near layered plasmonic structures, we study the simple situations of a QE above a bare and dielectric-covered metal substrate for which analytical results exist\cite{ford}. Moreover, we choose the emission wavelength of QE to be $\lambda=780$\,nm, metal is assumed to be gold, and the 50\,nm thick dielectric spacer represents silicon dioxide (SiO$_2$). In the case of QE in close vicinity of a bare gold film (Figures \ref{fig:DipolesLayeredGeom}a and \ref{fig:DipolesLayeredGeom}b), one notices the different dependencies of decay channels as a function of separation. For nanometer separations, the total decay rate is strongly enhanced, with the absorption in gold being the dominant decay channel, i.e. quenching of QE. At intermediate distances ($20$\,nm$<z_0<300$\,nm), however, vertical and horizontal QEs preferentially decay into SPPs and freely propagating waves, respectively. Finally, for large separations ($z_0>300$\,nm) the total decay rate approaches the free-space value, although the dominant radiation channel shows oscillatory behavior related to the interference with the reflected light. Note also that, unlike other studies\cite{ford}, we attribute power lost upon emitted light on reflection to the non-radiative decay channel, which is why this decay mechanism for $z_0\rightarrow\infty$ decreases to a constant value (marked with dotted-line in Figure \ref{fig:DipolesLayeredGeom}) and not zero as the SPP contribution. In regard to numerical calculations (markers in Figure \ref{fig:DipolesLayeredGeom}), we see an almost perfect overlap with analytical results for the non-negligible decay channels. It is only for large separations, when the relative decay rates into SPPs and Ohmic losses are of the order $\sim 10^{-2}$, that these calculations deviate from analytical curves. We ascribe this limited accuracy on the second decimal to factors like truncation of simulation domain, meshing, and two successive numerical surface integrations (eqs \ref{eq:gammarad} and \ref{eq:E2}).

In the second case, consisting of a QE above a gold substrate overlaid with 50\,nm of SiO$_2$ (Figures \ref{fig:DipolesLayeredGeom}c and \ref{fig:DipolesLayeredGeom}d), we benchmark the implementation for the more complicated three-layer geometry. Since the QE for all $z_0$ is kept at a distance to the metal interface, the total decay rates are only moderately increased, with a weak dependence on the separation distance for $z_0<100$\,nm. In fact, the similar optical properties of SiO$_2$ and air (compared to that of gold) result in QE decay rate dependencies on separation that qualitatively resemble those from a bare gold substrate when QE is displaced by $\sim 70$\,nm, corresponding to the optical path in the SiO$_2$ spacer. Accounting for this QE displacement between the two systems, it is seen that for small separations the spacer promotes the probability of QE relaxation by SPP excitation. %Interestingly, the vertical QE mainly decays into SPP, whereas the horizontal QE shows non-negligible contributions to the radiative and SPP decay rates. As in the previous case, for $z_0\sim\lambda$ we approach the situation of QE in homogeneous environment where only radiative decay can occur (assuming intrinsic quantum yield of one).
Importantly, the numerical calculations again show good agreement with analytical results for the dominant decay rates, illustrating only noticeable deviation for the small absorptive decay rate and, for large separations, the SPP contribution.

\subsection{Quantum Emitter near Gap-Plasmon Resonator}
Gap-surface plasmon resonators considered in this work consist of a gold nanobrick of height $t$ and width $w$ on top of a SiO$_2$ spacer of thickness $t_s$ and optically thick gold film (Figure \ref{fig:GSPresonator}a). Such metal-insulator-metal configurations behave as Fabry-Perot resonators in which plasmonic resonances correspond to standing waves of gap-surface plasmon (GSP) modes, arising due to the efficient reflection of GSP modes at nanobrick terminations\cite{nielsen}. Accordingly, GSP-resonators show easy scaling of resonance wavelengths by either the nanobrick width or gap thickness\cite{leveque,jung2}, while the intrinsic properties of GSP modes allow for high field enhancement in the gap region\cite{nielsen} and configurations that either efficiently scatter or absorb light\cite{pors} (facilitating, e.g., the design of black\cite{nielsen2} and colored\cite{roberts} metasurfaces), or launch SPPs\cite{liu,pors2}. It should be noted that the methodology proposed here can easily be extended to quantitatively study the optical characteristics of GSP-resonators, which include not only scattering and absorption cross sections but also a SPP cross section, describing the effective size of a resonator with respect to launching SPPs (see Methods section). As instructive examples, Figures \ref{fig:GSPresonator}b and \ref{fig:GSPresonator}c display cross sections normalized to the geometrical area $w^2$ for two different configurations, both showing the fundamental GSP-resonance at $\lambda=780$\,nm. It is clear that at resonance both GSP-resonators efficiently interact with the normal incident $x$-polarized light, demonstrating extinction cross sections of $\sim \times 35$ larger than the geometrical area. In the first case (Figure \ref{fig:GSPresonator}b), due to the relatively weakly confined GSP mode, the localized plasmon decays mainly via reradiation into free space ($\sim 50$\%), followed by SPP ($\sim 30$\%) and absorption ($\sim 20$\%). By decreasing the gap thickness $t_s$ the GSP mode becomes increasingly confined to the spacer region below the nanobrick, hence reducing scattering loss by reflection at end terminations at the expense of increased absorption. This is clearly seen for $t_s=20$\,nm (Figure \ref{fig:GSPresonator}c) where absorption loss ($\sim 45$\%) exceeds the scattering contribution ($\sim 40$\%), with the SPP decay channel ($\sim 15$\%) playing a minor role. Note also a reduced line width of the GSP resonance for $t_s=20$\,nm, which owes to a decrease in the electric dipole moment for decreasing spacer thickness in favor of a magnetic response, decreasing thereby the radiative damping\cite{pors}. The strong localized electric field below the nanobrick at GSP resonance is exemplified in insets of Figures \ref{fig:GSPresonator}b and \ref{fig:GSPresonator}c, illustrating the general features of the fundamental GSP resonance with zero field in the center, maximum field below the rim of the nanobrick, dominantly $z$-directed electric field, and increasing field enhancement for decreasing spacer thickness.

\begin{figure}[H]
	\centering
		\includegraphics[width=0.85\linewidth]{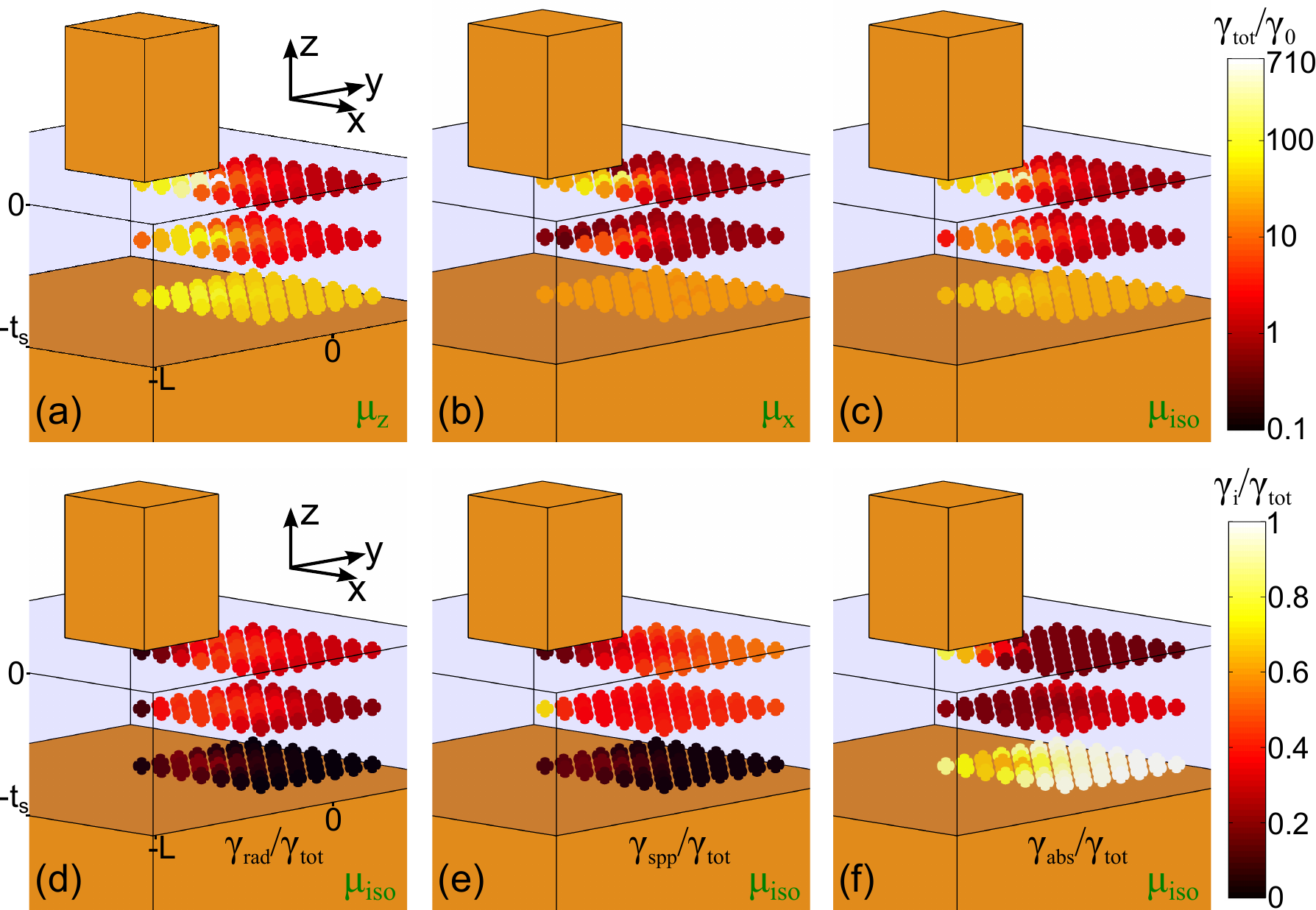}
	\caption{Total decay rate modification for (a) $z$-directed, (b) $x$-directed, and (c) isotropically-oriented QE as a function of position in the dielectric spacer $\big[(x,y)\in [0;180]$\,nm in steps of $30$\,nm and $z=-5,-25,-45$\,nm$\big]$ of GSP resonator ($t=t_s=50$\,nm, $w=120$\,nm) at $\lambda=780$\,nm. (d-f) Distribution into the three decay channels as a function of position for isotropically-oriented QE. Note that the system is for ease of visualization not drawn to scale; domain parameter $L=250$\,nm.}
	\label{fig:EmitterNearGSPresonator}
\end{figure}

As the localized enhancement of the electric field near GSP resonators indicates a strongly position-dependent and increased local density of states, we proceed with a quantitative study of decay rate modifications, including distribution in the three decay channels, of QEs positioned in the spacer layer in close proximity of the resonant GSP resonator from Figure \ref{fig:GSPresonator}b with emission wavelength $\lambda=780$\,nm (see Figure \ref{fig:EmitterNearGSPresonator}; a two-dimensional representation of the data can be found in Figures S5 and S6 in SI). Moreover, we restrict (due to symmetry reasons) the calculations to the first quadrant in the $xy$-plane and three $z$-planes, while focusing on $z$-, $x$-, and isotropically-oriented QEs. The latter case is important in relation to experimental situations where the orientation of QEs is generally not known. Finally, one should note that decay rate modifications of $y$-directed QEs can be obtained from the results of $x$-directed counterparts by mirroring the data in the plane defined by the $z$-axis and the line $y=x$. Regarding Figures \ref{fig:EmitterNearGSPresonator}a-c, it is clear that for $5$\,nm separation between QE and gold film or nanobrick the total decay rates are enhanced by up to two orders of magnitude, with the largest increase observed for $z$-directed QEs due to the overlap of this orientation with the electric field of the GSP resonance. As further discussed below, one should take notice of the pronounced difference in contributions of decay channels for QEs close to gold film and nanobrick (Figures \ref{fig:EmitterNearGSPresonator}d-f). Furthermore, in relation to the inset in Figure \ref{fig:GSPresonator}b, one clearly observes the direct correlation between the spatial distribution of the enhanced electric field and the position-dependent total decay rate of $z$-directed QE, featuring the highest $\gamma_{\mathrm{tot}}$ at shortest distance to the vertex of the nanobrick (Figure \ref{fig:EmitterNearGSPresonator}a). The modification of total decay rate for $x$-directed QEs away from metal surfaces is, on the other hand, controlled by the smaller $x$-component of the GSP field; a fact seen in the center plane of Figure \ref{fig:EmitterNearGSPresonator}b where increased $\gamma_{\mathrm{tot}}$ occurs at and outside the rim of the nanobrick but not beneath it, as here metal boundary conditions enforce the electric field to be perpendicular to the gold surfaces.

Proceeding with the influence of different decay channels as a function of QE position in the spacer layer, we limit the discussion to the experimentally important isotropically-oriented QEs (Figures \ref{fig:EmitterNearGSPresonator}d-f), though calculations for $z$- and $x$-directed QEs can be found in Figures S7 and S8 in SI. Figures \ref{fig:EmitterNearGSPresonator}d-f display the probability of which a QE decays either radiatively (i.e., quantum yield), via SPPs, or non-radiatively. As expected from the study of the bare gold interface (Figure \ref{fig:DipolesLayeredGeom}), QEs mainly decay non-radiatively for nanometer separations to the gold film. On the other hand, it is evident that a QE positioned in the upper half of the spacer and in the neighborhood of the nanobrick may show moderate enhancement of the total decay rate (up to $\sim 50$, c.f. Figure \ref{fig:EmitterNearGSPresonator}c), with radiative and SPP efficiency of $\sim 0.4$ (Figures \ref{fig:EmitterNearGSPresonator}d and \ref{fig:EmitterNearGSPresonator}e). Importantly, even a QE positioned 5\,nm below the vertex of the nanobrick, featuring strongly enhanced total decay rate, displays quantum yield of $\sim 0.4$. Accordingly, QEs near GSP resonators may show strong enhancement of spontaneous emissions or function as an efficient local source for generation of SPPs. It should be noted that further enhancement of spontaneous emission can be achieved by reducing the thickness of the spacer and replacing gold with silver, as recently confirmed experimentally\cite{russell,rose}. As an example of the effect of spacer thickness reduction, QEs beneath the GSP resonator in Figure \ref{fig:GSPresonator}c shows up to three orders of magnitude increase in the total decay rate at $\lambda=780$\,nm, with isotropically-oriented QEs still featuring quantum yields of $\sim 0.4$ in the center of the spacer layer (Figures S9-S12 in SI).

As QEs near metallic nanostructures excite the associated plasmonic resonances, which in turn strongly modifies the spontaneous decay rate, the direction of which photons are emitted into the far-field and the directional emission of SPPs will also be affected. We now study emission patterns of photons and SPPs for $z$- and $x$-directed QEs as a function of position in the spacer of GSP resonator in Figure \ref{fig:GSPresonator}b; emission patterns for $y$-oriented QEs can be found in Figure S13 in SI. To be specific, we limit the discussion to QEs in the center of the spacer and moving along the $x$-axis (Figure \ref{fig:RadSPPpattern}a), which entails modification in the radiation patterns only in the $xz$-plane (Figure \ref{fig:RadSPPpattern}c). Likewise, the configuration ensures that SPP emission patterns are mirror symmetric with the $xz$-plane, hence allowing us to only display the positive half of the $xy$-plane (Figure \ref{fig:RadSPPpattern}d). Furthermore, note that all emission patterns are normalized to unity, whereas the efficiency of emission can be evaluated from Figure \ref{fig:RadSPPpattern}b, displaying the relative decay rates as a function of $x$-coordinate. Regarding radiation patterns (Figure \ref{fig:RadSPPpattern}c), it is clear (and expected) that the symmetric configuration ($x=0$\,nm) results in emission equivalent to a QE above a gold substrate without the presence of the nanobrick. However, as soon as the symmetry is broken, the GSP mode is excited and, thus, leads to radiation patterns mainly dictated by this mode (see Figure S14 in SI). Interestingly, note how the angle of the main lobe of radiation for $x$-directed QE is especially sensitive to the position and that $z$- and $x$-directed QEs display main lobe radiation on opposite sides of the surface normal ($z$-axis) -- the latter observation indicates a possible route to experimentally determine the orientation of QE. Finally, it is worth noting that despite the minimal influence of the GSP resonator on QE decay rates for large separations ($x>150$\,nm), the radiation patterns are still strongly modified, demonstrating for $x=250$\,nm multi-lobe behavior as a result of interference between direct QE radiation and radiation emanating from the excitation of the GSP mode. With respect to excitation of SPPs (Figure \ref{fig:RadSPPpattern}d), it is clear that displacing the QE along the $x$-direction leads to noticeable unidirectional excitation of SPPs. For example, for the efficient $z$-directed QE at $x=30$\,nm ($\gamma_{\mathrm{spp}}/\gamma_0\sim 15$, c.f. Figure \ref{fig:RadSPPpattern}b) the ratio between power flowing along the $\pm x$-direction is $D \sim 10$. Moreover, one notices that for a large range of $x$-positions ($x=30-150$\,nm) SPPs are predominantly launched along $\pm x$-direction (i.e., $D \gtrless 1$) for $z$- and $x$-directed QEs, respectively.% (for $x=30-150$\,nm) $D \gtrless 1$ for $z$- and $x$-directed QEs, respectively.%; an observation that in experimental situations may give insight into the orientation of QE by utilizing a leakage radiation microscopy setup.     

\section{CONCLUSIONS}
In summary, we propose a numerical framework to accurately calculate radiative, SPP and non-radiative decay rates for QEs close to metal films and arbitrarily-shaped nanostructures. The methodology is based on Huygen's equivalence principle and knowledge of the angular spectrum Green's dyadics for the layered reference system, allowing one to accurately calculate the electric far-field. Using the finite-element approach, we verify the calculation procedure for QEs above a gold substrate, demonstrating consistency with analytical results for the dominant decay channels. As GSP-resonators recently have shown the ability to strongly enhance spontaneous emission of QEs embedded in the spacer layer\cite{russell,rose,ciraci}, we discuss such a configuration in detail, emphasizing the strong position and orientation dependencies of the three decay channels, radiation patterns, and directionality of SPP excitation -- all features that can be of interest depending on the application. 

We would like to emphasize that the methodology can be extended to systems of arbitrary number of layers and optically-thin metal films, with the possibility of constructing the associated Green's dyadics in a rather simple way, as outlined in the text. Moreover, the approach enables one to study the delicate interplay between geometric parameters, material properties, and QE position and orientation in relation to the total decay rate and significance of decay channels, hereby allowing not only to optimize for strong spontaneous emission but alternatively for efficient (and strongly directional) excitation of SPPs, which is of great interest for developing compact plasmonic circuitry.

\section{METHODS}

\subsection{Finite Element Modeling}
All calculations have been performed using the commercial finite element software Comsol Multiphysics (ver. 4.4) in which the simulation domain is truncated using manually-implemented unidirectional perfectly matched layers. Quantum emitters are represented by electric point dipoles that mathematically are equivalent to a small line current, described by a constant product of current and length, in the limit of vanishing length. It should be noted that the calculation of the electric far-field, as defined in eqs \ref{eq:E2} and \ref{eq:Efar} (featuring two sets of spatial coordinates), can be directly implemented in Comsol using the \emph{dest}-operator.

\begin{figure}[H]
	\centering
		\includegraphics[width=0.4\linewidth]{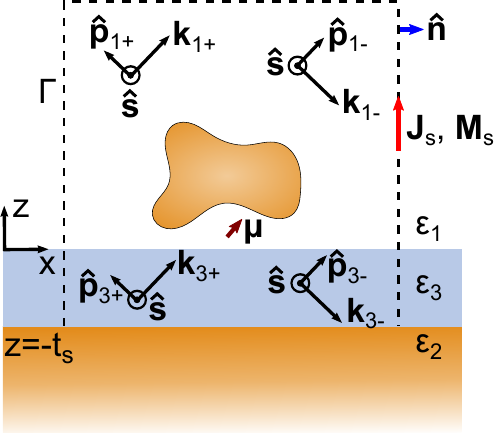}
	\caption{Sketch of plasmonic layered system, consisting of an optically thick metal film, a sub-wavelength dielectric spacer, and a quantum emitter in close vicinity of a arbitrary-shaped metallic nanostructure.}
	\label{fig:Sketch3LayerSystem}
\end{figure}

\begin{figure}[H]
	\centering
		\includegraphics[width=0.4\linewidth]{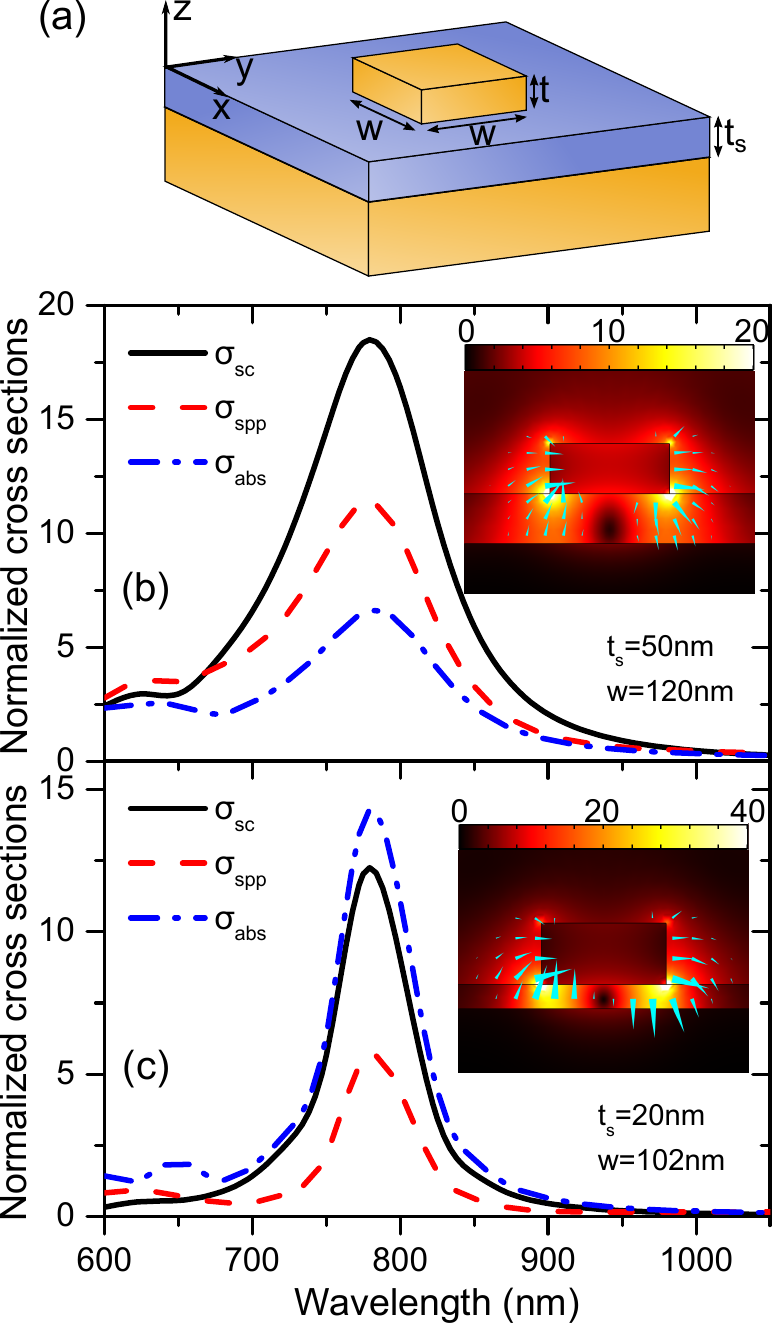}
	\caption{(a) Sketch of gap-surface plasmon resonator, defined by optically-thick metal substrate, dielectric spacer thickness $t_s$, and nanobrick height $t$ and width $w$. Normalized absorption, scattering, and SPP cross sections for GSP-resonator of gold and SiO$_2$ for (b) $t=t_s=50$\,nm and $w=120$\,nm, and (c) $t=50$\,nm, $t_s=20$\,nm and $w=102$\,nm. The incident $x$-polarized plane wave propagates normal to the surface. Insets in b,c display the electric field enhancement in the $xz$-plane at $\lambda=780$\,nm, with cones illustrating the direction of the field. Note that the scale of the color bar is chosen to emphasize the field distribution below the nanobrick rather than the strongly enhanced field at nanobrick corners. }
	\label{fig:GSPresonator}
\end{figure}

Regarding the calculation of optical cross-sections for GSP-resonators (Figure \ref{fig:GSPresonator}), the approach is based on the division of the total electric field in medium 1 and 3 (see Figure \ref{fig:Sketch3LayerSystem}) in two parts: $\mathbf{E}=\mathbf{E}_{\mathrm{ref}}+\mathbf{E}_{\mathrm{sc}}$, where $\mathbf{E}_{\mathrm{ref}}$ is the reference field (i.e., without nanobrick) and $\mathbf{E}_{\mathrm{sc}}$ is the scattered field present due to the nanobrick. 
%$\mathbf{E}=\mathbf{E}_{\mathrm{i}}+\mathbf{E}_\mathrm{r}+\mathbf{E}_{\mathrm{sc}}$, where $\mathbf{E}_{\mathrm{i}}$ is the incident electric field, $\mathbf{E}_{\mathrm{r}}$ is the specularly reflected field from the layered geometry without the nanobrick (known analytically), and $\mathbf{E}_{\mathrm{sc}}$ is the scattered field present due to the nanobrick.
It is only the latter quantity of the electromagnetic field (i.e., $\mathbf{E}_{\mathrm{sc}}$ and $\mathbf{H}_{\mathrm{sc}}$) that defines the equivalent electric and magnetic surface currents
\begin{equation}
\mathbf{J}_s=\hat{\mathbf{n}}\times\mathbf{H}_{\mathrm{sc}} \ \ \ ,\ \ \ \mathbf{M}_s=-\hat{\mathbf{n}}\times\mathbf{E}_{\mathrm{sc}},
\label{eq:sc2}
\end{equation}
which are used in calculating the electric far-field ($\mathbf{E}_{\mathrm{ff}}$) and, hence, the scattering and SPP cross-sections:
\begin{align}
& \sigma_{\mathrm{sc}}(\omega)=\frac{1}{2\eta_1I_i}\iint_{\mathrm{HS}}{|\mathbf{E}_{\mathrm{ff}}|^2}\,dS, \\
& \sigma_{\mathrm{spp}}(\omega)=\frac{1}{I_i}\iint_\Gamma{\mathbf{S}_{\mathrm{sc}}\cdot\hat{\mathbf{n}}}\,dS-\sigma_{\mathrm{sc}}.
\end{align}
Here, $I_i=E_0^2/(2\eta_1)$ is the intensity of the incident plane wave with amplitude $E_0$ and wave impedance $\eta_1$, $\mathbf{S}_\mathrm{sc}=1/2\mathrm{Re}\left\{\mathbf{E}_{\mathrm{sc}}\times\mathbf{H}_\mathrm{sc}^*\right\}$ is the time-averaged scattered Poynting vector, and $\hat{\mathbf{n}}$ is the outward-pointing unit vector to the surface $\Gamma$ as defined in Figure \ref{fig:Sketch3LayerSystem}. The absorption cross-section is defined as 
\begin{equation}
\sigma_\mathrm{abs}(\omega)=-\frac{1}{I_i}\iint_\Gamma{\left(\mathbf{S}-\mathbf{S}_\mathrm{ref}\right)\cdot\hat{\mathbf{n}}}\,dS,
\end{equation}
where $\mathbf{S}$ and $\mathbf{S}_\mathrm{ref}$ are the Poynting vectors for the total and reference field, respectively. The negative sign in front of the integral ensures a positive quantity and indicates that power is lost inside the surface $\Gamma$ due to the presence of the nanobrick. It should be noted that gold permittivity is described by interpolated experimental data\cite{johnson} while the spacer of silicon dioxide takes on the constant refractive index of $1.45$.

\begin{figure}[H]
	\centering
		\includegraphics[width=8.6cm]{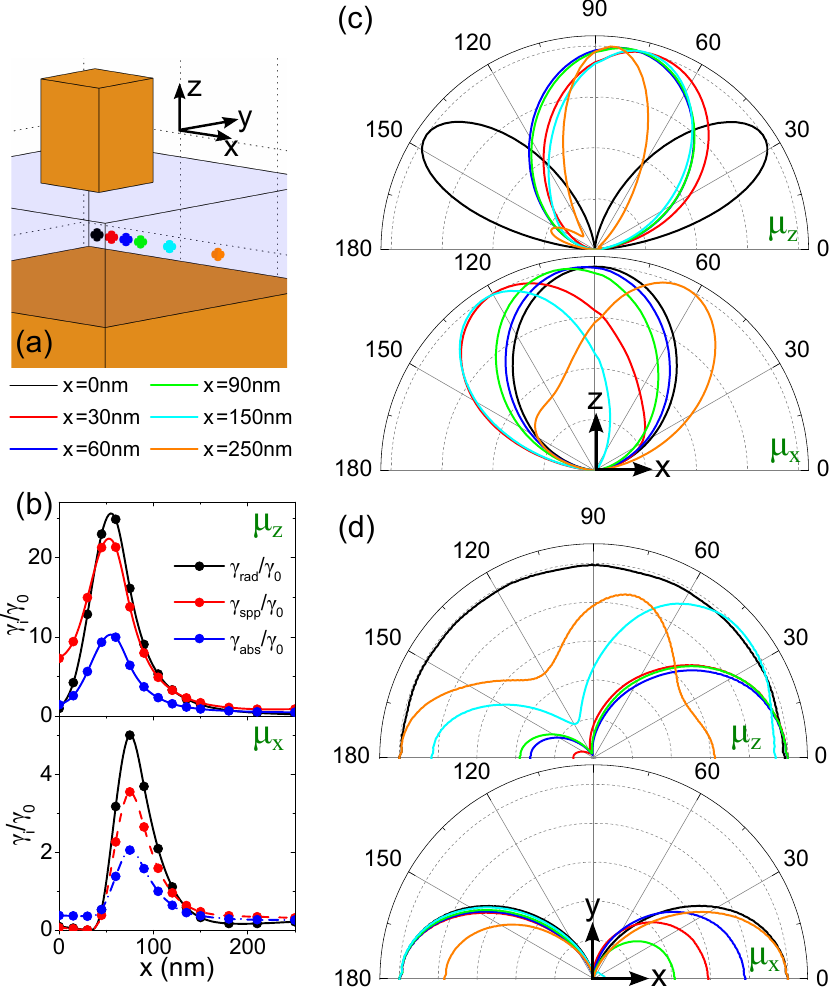}
	\caption{(a) Sketch of GSP-resonator for $t=t_s=50$\,nm and $w=120$\,nm, and indications of QE positions for varying $x$-coordinate, $y=0$\,nm, and $z=-25$\,nm. (b) Relative decay rates for $z$- and $x$-directed QE as a function of $x$-coordinate. Normalized (c) radiation patterns in the $xz$-plane and (d) SPP patterns in the $xy$-plane for the six QE positions indicated in panel a.}
	\label{fig:RadSPPpattern}
\end{figure}

The radiation patterns in Figure \ref{fig:RadSPPpattern}b are obtained by plotting $|\mathbf{E}_{\mathrm{ff}}|^2$, whereas the SPP emission patterns in Figure \ref{fig:RadSPPpattern}d result from plotting the squared norm of the  normal component of the electric field at the surface of the gold film on a circle of radius $r=2\lambda=1.56$\,$\mu$m\cite{sondergaard3}.  
%obtained by integrating the dissipated power density,$Q_h(\mathbf{r})=\omega \mathrm{Im}\{\varepsilon_{au}\}|\mathbf{E}|^2/2$, in the nanobrick and the part of gold film inside $\Gamma$, subtracting the absorbed power in the gold film from the reference geometry (i.e., without nanobrick), and normalizing with the intensity of the incident plane wave. Additionally, gold permittivity is described by interpolated experimental data,  mention dest-operator, SPP cross section, how to calculate SPP patterns, 

%% classes.  Rather than use \section, an appropriate macro is
%% provided that will always work.
%%%%%%%%%%%%%%%%%%%%%%%%%%%%%%%%%%%%%%%%%%%%%%%%%%%%%%%%%%%%%%%%%%%%%
%\acknowledgement
\begin{acknowledgement}
We acknowledge financial support for this work from the Danish Council for Independent Research (the FNU project, contract no. 12-124690) and European Research Council, Grant 341054 (PLAQNAP). 
\end{acknowledgement}

%% \suppinfo macro.
%%%%%%%%%%%%%%%%%%%%%%%%%%%%%%%%%%%%%%%%%%%%%%%%%%%%%%%%%%%%%%%%%%%%%
%\suppinfo
%
%The entire \textsf{achemso} bundle is generated by running
%\texttt{achemso.dtx} through \TeX. Running \LaTeX\ on the same file
%will generate the general documentation for both the class and
%package files.

\suppinfo 
Calculation of the dependence of decay rates on the size of the surface $\Gamma$ for vertical QE in air and positioned 20\,nm above a gold substrate; calculation of decay rates for vertical QE in air with emission wavelength of 500\,nm and 550\,nm as a function of distance to a gold substrate; direct and indirect (see eq \ref{eq:gammaabs}) calculations of the non-radiative decay rate for vertical QE in air with varying distance to a gold substrate; calculation of radiation pattern of vertical QE in air positioned 20\,nm above a gold substrate for a closed and open surface $\Gamma$; calculation of position-dependent decay rate distributions for $z$- and $x$-directed QEs situated in the spacer of GSP-resonator with gap thickness $t_s=50$\,nm; position- and orientation-dependence of total decay rate and probability of relaxation paths for QE in the spacer of GSP-resonator with $t_s=20$\,nm; radiation patterns and directionality of SPP emission for $y$-directed QE as a function of $x$-coordinate in the center of spacer of GSP-resonator ($t_s=50$\,nm); radiation pattern of the fundamental resonance of GSP-resonators. 
%\bibliography{References_EmitterNearGSPresonator}

\providecommand*\mcitethebibliography{\thebibliography}
\csname @ifundefined\endcsname{endmcitethebibliography}
  {\let\endmcitethebibliography\endthebibliography}{}

%% graphical table of contents.
%%%%%%%%%%%%%%%%%%%%%%%%%%%%%%%%%%%%%%%%%%%%%%%%%%%%%%%%%%%%%%%%%%%%%

\newpage

\begin{tocentry}

\centering\includegraphics[width=7.9cm]{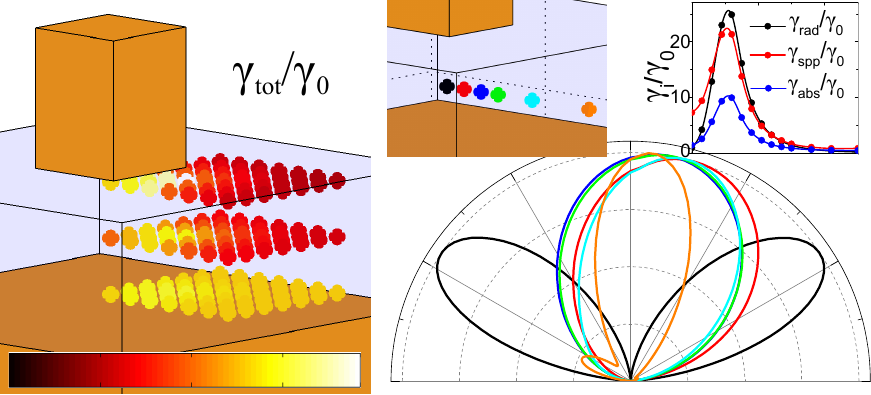}

\end{tocentry}

%% FIGURES %%

\pagebreak

\pagebreak 

\pagebreak

\pagebreak

\pagebreak

\end{document}